\documentclass[12pt,letterpaper]{article}

\usepackage{natbib,epsfig,graphicx,setspace}
\usepackage{amsmath,amsthm,amssymb,enumerate}
\usepackage{mathrsfs}
\usepackage{bm,verbatim,subfigure}
\usepackage{booktabs}
\usepackage{color, xcolor}
\usepackage{booktabs}
\usepackage{rotating}

\newtheorem{thm}{Theorem}[section]

\def\bsigma{{\mbox{\boldmath $\sigma$}}}
\def\btheta{{\mbox{\boldmath $\theta$}}}
\def\bbeta{{\mbox{\boldmath $\beta$}}}
\def\balpha{{\mbox{\boldmath $\alpha$}}}
\def\bpi{{\mbox{\boldmath $\pi$}}}

\def\bS{{\mbox{\boldmath $S$}}}
\def\bm{{\mbox{\boldmath $m$}}}
\def\bx{{\mbox{\boldmath $x$}}}

\def\bW{{\mbox{\boldmath $W$}}}
\def\bA{{\mbox{\boldmath $A$}}}
\def\bQ{{\mbox{\boldmath $Q$}}}
\def\bR{{\mbox{\boldmath $R$}}}
\def\b0{{\mbox{\boldmath $0$}}}
\def\by{{\mbox{\boldmath $y$}}}
\def\bta{{\mbox{\boldmath $\eta$}}}
\def\blambda{{\mbox{\boldmath $\lambda$}}}
\def\bLambda{{\mbox{\boldmath $\Lambda$}}}
\def\bI{{\mbox{\boldmath $I$}}}

\numberwithin{equation}{section} 

\numberwithin{equation}{section} 

\def\boxit#1{\vbox{\hrule\hbox{\vrule\kern6pt\vbox{\kern6pt#1\kern6pt}\kern6pt\vrule}\hrule}}

\begin{document}
\title{Mixture of Regression Models with Single-index}

\author{Sijia Xiang \thanks{Corresponding Author,
School of Mathematics and Statistics, Zhejiang University of Finance and Economics. E-mail:
sjxiang@zufe.edu.cn. Xiang's research is supported by Zhejiang Provincial NSF of China grant LQ16A010002.}, and Weixin Yao
\thanks{Department of Statistics, University of California, Riverside. E-mail: weixin.yao@ucr.edu. Yao's research is supported by NSF grant DMS-1461677.}}

\date{}
\maketitle{}
\begin{abstract}
In this article, we propose a class of semiparametric mixture regression models with single-index. We argue that many recently proposed semiparametric/nonparametric mixture regression models can be considered as special cases of the proposed model. However, unlike existing semiparametric mixture regression models, the new proposed model can easily incorporate multivariate predictors into the nonparametric components. Backfitting estimates and the corresponding algorithms have been proposed to achieve the optimal convergence rate for both the parameters and the nonparametric functions. We show that nonparametric functions can be estimated with the same asymptotic accuracy as if the parameters were known and the index parameters can be estimated with the traditional parametric root $n$ convergence rate. Simulation studies and an application of NBA data have been conducted to demonstrate the finite sample performance of the proposed models.
\end{abstract}

\vskip 20pt


\section{Introduction}

Mixtures of regression models are commonly used to reveal the relationship among interested variables if the whole population is not homogeneous and consists of several homogeneous subgroups. It has been widely used in many areas such as econometrics, biology, and epidemiology. Recently, many semiparametric mixture models have been proposed. See, for example, Young and Hunter (2010), Huang and Yao (2012), Huang et al. (2013), Cao and Yao (2012), Xiang and Yao (2015), among others.

In this article, we apply the idea of single-index model to mixture of regression models, and propose a mixture of single-index models (MSIM) and a mixture of regression models with varying single-index proportions (MRSIP). Huang et al. (2013) proposed the nonparametric mixture of regression models $Y|_{X=x}\sim \sum_{j=1}^k \pi_j(x)\phi(Y_i|m_j(x),\sigma_j^2(x))$, and developed an estimation procedure by employing kernel regression. However, the above model is not very applicable to multivariate predictors due to the so called ``curse of dimensionality". The proposed mixture of single-index models can naturally incorporate the multivariate predictors and relax the traditional parametric assumption of mixture of regression models.

In some cases, we might want to assume linearity in the mean functions. Therefore, the proposed MRSIP keeps the easy interpretation of the linear component regression functions while assuming that the mixing proportions are smooth functions of an index $\balpha^T\bx$.

We show the identifiability of each model under some regularity conditions. To achieve the optimal convergence rate for the global parameters and nonparametric functions, we propose backfitting estimates using the kernel regression technique. We have shown that the nonparametric functions can be estimated with the same rate as if the parameters were known, and the parameters can be estimated with the same rate of convergence, $n^{-1/2}$, that is achieved in a parametric model. Numerical studies are used to demonstrate the effectiveness of the proposed new models, and we discuss the selection of the two models in the real data analysis.

The rest of the paper is organized as follows. In Section \ref{sec:p3-s2}, we introduce the MSIM and study its identifiability result. A one-step estimate and a fully-iterated backfitting estimate have been proposed, and their asymptotic properties are studied. Section \ref{sec:p3-s3} discusses the MRSIP and its identifiability. A fully-iterated estimate and its asymptotic properties are also studied. In Section \ref{sec:p3-s4}, we use Monte Carlo studies and a real data example to demonstrate the finite sample performance of the proposed estimates. A discussion section ends the paper.

\section{Mixture of Single-index Models (MSIM)}
\label{sec:p3-s2}
\subsection{Model Definition and Identifiability}
Assume that $\{(\bx_i,Y_i),i=1,...,n\}$ is a random sample from population $(\bx,Y)$. Throughout this article, we assume that $\bx$ is $p$-dimensional and $Y$ is univariate. Let $\mathcal{C}$ be a latent variable, and we assume that conditional on $\bx$, $\mathcal{C}$ has a discrete distribution $P(\mathcal{C}=j|\bx)=\pi_j(\balpha^T\bx)$ for $j=1,...,k$. Conditional on $\mathcal{C}=j$ and $\bx$, $Y$ follows a normal distribution with mean $m_j(\balpha^T\bx)$ and variance $\sigma_j^2(\balpha^T\bx)$. We assume that $\pi_j(\cdot)$, $m_j(\cdot)$, and $\sigma_j^2(\cdot)$ are unknown but smooth functions, and therefore, without observing $\mathcal{C}$, the conditional distribution of $Y$ given $\bx$ can be written as:
\begin{equation}
Y|_{\bx}\sim\sum_{j=1}^k\pi_j(\balpha^T\bx)\phi(Y_i|m_j(\balpha^T\bx),\sigma_j^2(\balpha^T\bx)),
\label{p3-model}
\end{equation}
where $\phi(y|\mu,\sigma^2)$ is the normal density with mean $\mu$ and variance $\sigma^2$. Throughout the paper, we assume that $k$ is fixed, and refer to model (\ref{p3-model}) as a finite semiparametric mixture of regression models, since $\pi_j(\cdot)$, $m_j(\cdot)$ and $\sigma_j^2(\cdot)$ are all nonparametric. When $k=1$ and $\pi_j(\cdot)$ and $\sigma_j^2(\cdot)$ are constant, model (\ref{p3-model}) reduces to a single index model (Ichimura, 1993; H$\ddot{a}$rdle et al., 1993). If $\pi_j(\cdot)$ and $\sigma_j^2(\cdot)$ are constant, and $m_j(\cdot)$ are identity functions, then model (\ref{p3-model}) reduces to a finite mixture of linear regression models (Goldfeld and Quandt, 1973). If $\bx$ is a scalar, then model (\ref{p3-model}) reduces to the nonparametric mixture of regression model proposed by Huang et al. (2013). Therefore, the proposed model (\ref{p3-model}) is a natural generalization of many existing popular models.

Compared to Huang et al. (2013), the appeal of the proposed MSIM is that by focusing on an index $\balpha^T\bx$, the so-called ``curse of dimensionality'' in fitting multivariate nonparametric regression functions is avoided. It is of dimension-reduction structure in the sense that, if we can estimate the index $\balpha$ efficiently, then we can use the univariate $\hat{\balpha}^T\bx$ as the covariate and simplify the model (\ref{p3-model}) to the nonparametric mixture regression model proposed by Huang et al. (2013), and thus avoid the curse of dimensionality when nonparametric smoothing is employed. Therefore, model (\ref{p3-model}) is a reasonable compromise between fully parametric and fully nonparametric modeling.


Identifiability is a major concern for most mixture models. Some well known results for identifiability of finite mixture models include: mixture of univariate normals is identifiable up to relabeling (Titterington et al. 1985) and finite mixture of regression models is identifiable up to relabeling provided that covariates have a certain level of variability (Henning, 2000). The following theorem gives the result on identifiability of model (\ref{p3-model}) and its proof is given in Section \ref{p3-sec:proofs}.

\begin{thm}
\label{p3-thm:iden}
Assume that
(i) $\pi_j(z)$, $m_j(z)$, and $\sigma_j^2(z)$ are differentiable and not constant on the support of $\balpha^T\bx$, $j=1,...,k$; (ii) The component of $\bx$ are continuously distributed random variables that have a joint probability density function; (iii) The support of $\bx$ is not contained in any proper linear subspace of $\mathbb{R}^p$; (iv) $\|\balpha\|=1$ and the first nonzero element of $\balpha$ is positive; (v) Any two curves $(m_i(z),\sigma_i^2(z))$ and $(m_j(z),\sigma_j^2(z))$, $i\neq j$, are transversal.
Then, model (\ref{p3-model}) is identifiable.
\end{thm}

The transversality of two smooth curves (Huang et al., 2013) implies that the mean and variance functions of any two components cannot be tangent to each other.

\subsection{Estimation Procedure and Asymptotic Properties}
\label{sec:p3-s3}
In this subsection, we propose a one-step estimate and a fully iterative backfitting estimate to achieve the optimal convergence rate for both the index parameter and nonparametric functions.

Let $\ell^{*(1)}(\bpi,\bm,\bsigma^2,\balpha)$ be the log-likelihood of the collected data $\{(\bx_i,Y_i),i=1,...,n\}$. That is:
\begin{equation}
\ell^{*(1)}(\bpi,\bm,\bsigma^2,\balpha)=\sum_{i=1}^n\log\{\sum_{j=1}^k\pi_j(\balpha^T\bx_i)\phi(Y_i|m_j(\balpha^T\bx_i),\sigma_j^2(\balpha^T\bx_i))\},
\label{p3-lstar}
\end{equation}
where $\bpi(\cdot)=\{\pi_1(\cdot),...,\pi_{k-1}(\cdot)\}^T$, $\bm(\cdot)=\{m_1(\cdot),...,m_k(\cdot)\}^T$, and $\bsigma^2(\cdot)=\{\sigma_1^2(\cdot),...,\sigma_k^2(\cdot)\}^T$. Since $\bpi(\cdot)$, $\bm(\cdot)$ and $\bsigma^2(\cdot)$ consist of nonparametric functions, (\ref{p3-lstar}) is not ready for maximization.

If $\hat{\balpha}$ is an estimate of $\balpha$, then $\bpi(\cdot)$, $\bm(\cdot)$ and $\bsigma^2(\cdot)$ can be estimated locally by maximizing the following local log-likelihood function:
\begin{equation}
\ell^{(1)}_1(\bpi,\bm,\bsigma^2)=\sum_{i=1}^n\log\{\sum_{j=1}^k\pi_j(\hat{\balpha}^T\bx_i)\phi(Y_i|m_j(\hat{\balpha}^T\bx_i),\sigma_j^2(\hat{\balpha}^T\bx_i))\}K_h(\hat{\balpha}^T\bx_i-z),
\label{p3-l1}
\end{equation}
where $K_h(z)=\frac{1}{h}K(\frac{z}{h})$ and $K(\cdot)$ is a kernel density function.

Let $\hat{\bpi}(\cdot)$, $\hat{\bm}(\cdot)$ and $\hat{\bsigma}^2(\cdot)$ be the result of maximizing (\ref{p3-l1}). We can then further update the estimate of $\balpha$ by maximizing
\begin{equation}
\ell^{(1)}_2(\balpha)=\sum_{i=1}^n\log\{\sum_{j=1}^k\hat{\pi}_j(\balpha^T\bx_i)\phi(Y_i|\hat{m}_j(\balpha^T\bx_i),\hat{\sigma}_j^2(\balpha^T\bx_i))\},
\label{p3-l2}
\end{equation}
with respect to $\balpha$.

\subsubsection{Computing Algorithm}
\label{sec:comp}
We now propose two effective algorithms to calculate the estimates.

\textbf{One-step Estimator (OS)}\\
\textbf{Step 1: Obtain an estimate of the index parameter $\balpha$}.\\
Apply sliced inverse regression (Li, 1991) to obtain the estimate of $\balpha$, denoted by $\hat{\balpha}$.\\
\textbf{Step 2: Modified EM-type algorithm to maximize $\ell^{(1)}_1$ in (\ref{p3-l1})}.\\
In Step 2, we propose a modified EM-type algorithm to maximize $\ell_1^{(1)}$ and obtain the estimators $\hat{\bpi}(\cdot)$, $\hat{\bm}(\cdot)$ and $\hat{\bsigma}^2(\cdot)$. In practice, we usually want to evaluate unknown functions at a set of grid points, which in this case, requires us to maximize local log-likelihood functions at a set of grid points. If we simply imply an EM algorithm, the labels in the EM algorithm may change at different grid points, and we may not be able to get smoothed estimated curves (Huang and Yao, 2012). Therefore, we propose the following modified EM-type algorithm, which estimates the nonparametric functions simultaneously at a set of grid points . Let $\{u_t,t=1,...,N\}$ be a set of grid points where some unknown functions are evaluated, and $N$ be the number of grid points.\\
\underline{\textbf{E-step:}}\\
Calculate the expectations of component labels based on estimates from $l^{th}$ iteration:
\begin{equation}
p_{ij}^{(l+1)}=\frac{\pi_j^{(l)}(\hat{\balpha}^T\bx_i)\phi(Y_i|m_j^{(l)}(\hat{\balpha}^T\bx_i),\sigma_j^{2(l)}(\hat{\balpha}^T\bx_i))}{\sum_{j=1}^k
\pi_j^{(l)}(\hat{\balpha}^T\bx_i)\phi(Y_i|m_j^{(l)}(\hat{\balpha}^T\bx_i),\sigma_j^{2(l)}(\hat{\balpha}^T\bx_i))}.
\label{p3-step1e}
\end{equation}
\underline{\textbf{M-step:}} \\
Update the estimates
\begin{align}
&\pi_j^{(l+1)}(z)=\frac{\sum_{i=1}^np_{ij}^{(l+1)}K_h(\hat{\balpha}^T\bx_i-z)}{\sum_{i=1}^nK_h(\hat{\balpha}^T\bx_i-z)}\label{p3-step1m1},\\
&m_j^{(l+1)}(z)=\frac{\sum_{i=1}^np_{ij}^{(l+1)}Y_iK_h(\hat{\balpha}^T\bx_i-z)}{\sum_{i=1}^np_{ij}^{(l+1)}K_h(\hat{\balpha}^T\bx_i-z)}\label{p3-step1m2},\\
&\sigma_j^{2(l+1)}(z)=\frac{\sum_{i=1}^np_{ij}^{(l+1)}(Y_i-m_j^{(l+1)}(z))^2K_h(\hat{\balpha}^T\bx_i-z)}{\sum_{i=1}^np_{ij}^{(l+1)}K_h(\hat{\balpha}^T\bx_i-z)},\label{p3-step1m3}
\end{align}
for $z\in\{u_t,t=1,...,N\}$. We then update $\pi_j^{(l+1)}(\hat{\balpha}^T\bx_i)$, $m_j^{(l+1)}(\hat{\balpha}^T\bx_i)$ and $\sigma_j^{2(l+1)}(\hat{\balpha}^T\bx_i)$, $i=1,...,n$ by linear interpolating $\pi_j^{(l+1)}(u_t)$, $m_j^{(l+1)}(u_t)$ and $\sigma_j^{2(l+1)}(u_t)$, $t=1,...,N$, respectively.

Note that in the M-step, the nonparametric functions are estimated simultaneously at a set of grid points, and therefore, the classification probabilities in the the E-step can be estimated globally to avoid the label switching problem (Yao and Lindsay, 2009).

\textbf{Fully Iterative Backfitting Estimator (FIB)}\\
To improve the estimation efficiency, we propose the following \emph{fully iterative backfitting estimator}.\\
\textbf{Step 1: Obtain an initial estimate of the index parameter $\balpha$}.\\
Apply sliced inverse regression to obtain an initial estimate of the index parameter $\balpha$, denoted by $\hat{\balpha}$.\\
\textbf{Step 2: Modified EM-type algorithm to maximize $\ell^{(1)}_1$ in (\ref{p3-l1})}.\\
With $\hat{\balpha}$, apply the modified EM-algorithm proposed above to obtain the estimators $\hat{\bpi}(\cdot)$, $\hat{\bm}(\cdot)$, and $\hat{\bsigma}^2(\cdot)$.\\
\textbf{Step 3: Updating the estimate of $\balpha$ by maximizing $\ell^{(1)}_2$ in (\ref{p3-l2})}.\\
Given $\hat{\bpi}(\cdot)$, $\hat{\bm}(\cdot)$, and $\hat{\bsigma}^2(\cdot)$ from Step 2, update the estimate of $\balpha$, denoted by $\hat{\balpha}$, which maximizes $\ell^{(1)}_2$ defined in (\ref{p3-l2}) using some numerical methods.\\
\textbf{Step 4: Iterate Step 2 - 3 until convergence}.

\subsubsection{Asymptotic Properties}

The asymptotic properties of the proposed estimates are investigated below.

Let $\btheta(z)=(\bpi^T(z),\bm^T(z),(\bsigma^2)^T(z))^T$. Define $\ell(\theta(z),y)=\log\sum_{j=1}^k\pi_j(z)\phi\{y|m_j(z),\sigma_j^2(z)\}$, $q_1(z)=\frac{\partial\ell(\theta(z),y)}{\partial\theta}$, $q_2(z)=\frac{\partial^2\ell(\theta(z),y)}{\partial\theta\partial\theta^T}$ and $\mathcal{I}^{(1)}_\theta(z)=-E[q_2(Z)|Z=z]$, $\Lambda_1(u|z)=E[q_1(z)|Z=u]$.

Under further conditions defined in Section \ref{p3-sec:proofs}, the properties of the one-step estimator when $\balpha$ is estimated to the order of $O_p(n^{-1/2})$ (i.e., at the usual parametric rate) is demonstrated in the following theorem.
\begin{thm}
\label{p3-thm:nonp}
Assume that conditions (C1)-(C7) in Section \ref{p3-sec:proofs} hold. Then, as $n\rightarrow\infty$, $h\rightarrow 0$ and $nh\rightarrow\infty$, we have
\begin{align}
\sqrt{nh}\{\hat{\btheta}(z)-\btheta(z)-\mathcal{B}_1+o_p(h^2)\}\overset{D}{\rightarrow} N\{0,\nu_0f^{-1}(z)\mathcal{I}^{(1)}_\theta(z)\},
\end{align}
where $\mathcal{B}_1(z)=\mathcal{I}^{(1)-1}_\theta\left\{\frac{f'(z)\Lambda_1^{'}(z|z)}{f(z)}+\frac{1}{2}\Lambda_1^{''}(z|z)\right\}\kappa_2h^2$, with $f(\cdot)$ the marginal density function of $\balpha^T\bx$, $\kappa_l=\int t^lK(t)dt$ and $\nu_l=\int t^lK^2(t)dt$.
\end{thm}

\noindent $\textbf{Remark 1.}$ The fully iterative backfitting estimator is at least as efficient as the one-step estimator, but the one-step estimator achieves the same efficiency in some important applications with added computational convenience. This information lower bound turns out to be the same as in Huang et al. (2013). Thus, the nonparametric functions can be estimated with the same rate of convergence as it would have if the one-dimension quantity $\balpha^T\bx$ were observable.

The next theorem shows that under further conditions, $\balpha$ can be estimated at the usual parametric rate using the fully iterated algorithm.
\begin{thm}
\label{p3-thm:index}
Assume that conditions (C1)-(C8) in Section \ref{p3-sec:proofs} hold. Then, as $n\rightarrow\infty$, $nh^4\rightarrow 0$, and $nh^2/\log(1/h)\rightarrow\infty$,
\begin{equation}
\sqrt{n}(\hat{\balpha}-\balpha)\overset{D}{\rightarrow} N(0,\bQ_1^{-1}),
\end{equation}
where $\bQ_1=E\left\{[\bx\btheta'(Z)]q_2(Z)[\bx\btheta'(Z)]^T\right\}-E\left[\bx\btheta'(Z)q_2(Z)\mathcal{I}^{(1)-1}_\theta(Z)E\{q_2(Z)[\bx\btheta'
(Z)]^T|Z\}\right]$.
\end{thm}

\section{Mixture of Regression Models with Varying Single-Index Proportions (MRSIP)}
\label{sec:p3-s3}
\subsection{Model Definition and Identifiability}
The MRSIP assumes that $P(\mathcal{C}=j|\bx)=\pi_j(\balpha^T\bx)$ for $j=1,...,k$, and conditional on $\mathcal{C}=j$ and $\bx$, $Y$ follows a normal distribution with mean $\bx^T\bbeta_j$ and variance $\sigma^2_j$. That is,
\begin{equation}
Y|_{\bx}\sim\sum_{j=1}^k\pi_j(\balpha^T\bx)N(\bx^T\bbeta_j,\sigma_j^2).
\label{p41-model}
\end{equation}
Since $\pi_j(\cdot)$'s are nonparametric, model (\ref{p41-model}) is also a finite semiparametric mixture of regression models.

\begin{thm}
\label{p4-thm:iden}
Assume that
(i) $\pi_j(z)>0$ are differentiable and not constant on the support of $\balpha^T\bx$, $j=1,...,k$; (ii) The component of $\bx$ are continuously distributed random variables that have a joint probability density function; (iii) The support of $\bx$ contains an open set in $\mathbb{R}^p$ and is not contained in any proper linear subspace of $\mathbb{R}^p$; (iv) $\|\balpha\|=1$ and the first nonzero element of $\balpha$ is positive; (v) $(\bbeta_j,\sigma_j^2)$, $j=1,...,k$, are distinct pairs.
Then, model (\ref{p41-model}) is identifiable.
\end{thm}

\subsection{Estimation Procedure and Asymptotic Properties}
The log-likelihood of the collected data is:
\begin{equation}
\ell^{*(2)}(\bpi,\bsigma^2,\balpha,\bbeta)=\sum_{i=1}^n\log\{\sum_{j=1}^k\pi_j(\balpha^T\bx_i)\phi(Y_i|\bx_i^T\bbeta_j,\sigma_j^2)\},
\label{p41-lstar}
\end{equation}
where $\bpi(\cdot)=\{\pi_1(\cdot),...,\pi_{k-1}(\cdot)\}^T$, $\bsigma^2=\{\sigma_1^2,...,\sigma_k^2\}^T$, and $\bbeta=\{\bbeta_1,...,\bbeta_k\}^T$. Since $\bpi(\cdot)$ consists of nonparametric functions, (\ref{p41-lstar}) is not ready for maximization.

If $(\hat{\balpha},\hat{\bbeta},\hat{\bsigma}^2)$ are estimates of $(\balpha,\bbeta,\bsigma^2)$, then $\bpi(\cdot)$ can be estimated locally by maximizing the following local log-likelihood function:
\begin{equation}
\ell^{(2)}_1(\bpi)=\sum_{i=1}^n\log\{\sum_{j=1}^k\pi_j(\hat{\balpha}^T\bx_i)\phi(Y_i|\bx_i^T\hat{\bbeta}_j,\hat{\sigma}^2_j)\}K_h(\hat{\balpha}^T\bx_i-z).
\label{p41-l1}
\end{equation}

Let $\hat{\bpi}(\cdot)$ be the result of maximizing (\ref{p41-l1}). We can then further update the estimate of $(\balpha,\bbeta,\bsigma^2)$ by maximizing
\begin{equation}
\ell^{(2)}_2(\balpha,\bbeta,\bsigma^2)=\sum_{i=1}^n\log\{\sum_{j=1}^k\hat{\pi}_j(\balpha^T\bx_i)\phi(Y_i|\bx_i^T\bbeta_j,\sigma^2_j)\}.
\label{p41-l2}
\end{equation}

\subsubsection{Computing Algorithm}

\textbf{Step 1: Obtain an initial estimate of $(\balpha,\bbeta,\bsigma^2)$}.\\
\textbf{Step 2: Modified EM-type algorithm to maximize $\ell^{(2)}_1$ in (\ref{p41-l1})}.\\
\underline{\textbf{E-step:}}\\
Calculate the expectations of component labels based on estimates from $l^{th}$ iteration:
\begin{equation}
p_{ij}^{(l+1)}=\frac{\pi^{(l)}_j(\hat{\balpha}^T\bx_i)\phi(Y_i|\bx_i^T\hat{\bbeta}_j,\hat{\sigma}_j^2)}
{\sum_{j=1}^k\pi^{(l)}_j(\hat{\balpha}^T\bx_i)\phi(Y_i|\bx_i^T\hat{\bbeta}_j,\hat{\sigma}_j^2)},j=1,...,k.
\end{equation}
\underline{\textbf{M-step:}}\\
Update the estimate
\begin{equation}
\pi_j^{(l+1)}(z)=\frac{\sum_{i=1}^np_{ij}^{(l+1)}K_h(\hat{\balpha}^T\bx_i-z)}{\sum_{i=1}^nK_h(\hat{\balpha}^T\bx_i-z)}
\end{equation}
for $z\in\{u_t,t=1,...,N\}$. We then update $\pi_j^{(l+1)}(\hat{\balpha}^T\bx_i)$, $i=1,...,n$ by linear interpolating $\pi_j^{(l+1)}(u_t)$, $t=1,...,N$.\\
\textbf{Step 3: Update $(\hat{\balpha},\hat{\bbeta},\hat{\bsigma}^2)$ by maximizing (\ref{p41-l2})}.\\
\textbf{Step 3.1: Given $\hat{\balpha}$, update $(\bbeta,\bsigma^2)$}.\\
\underline{\textbf{E-step:}}\\
Calculate the expectations of component identities:
\begin{equation}
p_{ij}^{(l+1)}=\frac{\hat{\pi}_j(\hat{\balpha}^T\bx_i)\phi(Y_i|\bx_i^T\bbeta^{(l)}_j,\sigma^{2(l)}_j)}
{\sum_{j=1}^k\hat{\pi}_j(\hat{\balpha}^T\bx_i)\phi(Y_i|\bx_i^T\bbeta^{(l)}_j,\sigma^{2(l)}_j)},j=1,...,k.
\end{equation}
\underline{\textbf{M-step:}}\\
Update $\bbeta$ and $\bsigma^2$:
\begin{align}
\bbeta_j^{(l+1)}&=(\bS^T\bR_j^{(l+1)}\bS)^{-1}\bS^T\bR_j^{(l+1)}\by,\\
\sigma_j^{2(l+1)}&=\frac{\sum_{i=1}^np_{ij}^{(l+1)}(Y_i-\bx_i^T\bbeta_j^{(l+1)})^2}{\sum_{i=1}^np_{ij}^{(l+1)}},
\end{align}
where $j=1,...,k$, $\bR_j^{(l+1)}=diag\{p_{ij}^{(l+1)},...,p_{nj}^{(l+1)}\}$, $\bS=(\bx_1,...,\bx_n)^T$.\\
\textbf{Step 3.2: Given $(\hat{\bbeta},\hat{\bsigma}^2)$, update $\balpha$.}\\
Given $(\hat{\bbeta},\hat{\bsigma}^2)$, maximize $\ell^{(2)}_3(\balpha)=\sum_{i=1}^n\log\{\sum_{j=1}^k\hat{\pi}_j(\balpha^T\bx_i)\phi(Y_i|\bx_i^T\hat{\bbeta}_j,\hat{\sigma}^2_j)\}$ to updates the estimate of $\balpha$, using some numerical methods.\\
\textbf{Step 3.3: Iterate Step 3.1-3.2 until convergence}.\\
\textbf{Step 4: Iterate Step 2-3 until convergence}.

\subsubsection{Asymptotic Properties}
Define $\bta=(\bbeta^T,(\bsigma^2)^T)^T$, $\blambda=(\balpha^T,\bta^T)^T$, and $\ell(\bpi(z),\blambda,\bx,y)=\log \sum_{j=1}^k\pi_j(z)\phi\{y|\bx^T\bbeta_j,\sigma_j^2\}$.
Let $q_{\pi}(z)=\frac{\partial\ell(\pi(z),\lambda,x,y)}{\partial\pi}$, $q_{\pi\pi}(z)=\frac{\partial^2\ell(\pi(z),\lambda,x,y)}{\partial\pi\partial\pi^T}$, and similarly, define $q_{\lambda}$, $q_{\lambda\lambda}$, and $q_{\pi\eta}$.
Denote $\mathcal{I}^{(2)}_\pi(z)=-E[q_{\pi\pi}(Z)|Z=z]$ and $\Lambda_2(u|z)=E[q_{\pi}(z)|Z=u]$.

Under further conditions, the properties of the estimator when $\blambda$ is estimated to the order of $O_p(n^{-1/2})$ (i.e., at the usual parametric rate) is demonstrated in the following theorem.
\begin{thm}
\label{p4-thm:nonp}
Assume that conditions (C1)-(C4) and (C9)-(C11) in Section \ref{p3-sec:proofs} hold. Then, as $n\rightarrow\infty$, $h\rightarrow 0$ and $nh\rightarrow\infty$, we have
\begin{align}
\sqrt{nh}\{\hat{\bpi}(z)-\bpi(z)-\mathcal{B}_2(z)+o_p(h^2)\}\overset{D}{\rightarrow} N\{0,\nu_0f^{-1}(z)\mathcal{I}^{(2)}_\pi(z)\},
\end{align}
where $\mathcal{B}_2(z)=\mathcal{I}_\pi^{(2)-1}\left\{\frac{f'(z)\Lambda_2'(z|z)}{f(z)}+\frac{1}{2}\Lambda_2''(z|z)\right\}\kappa_2h^2$.
\end{thm}

\begin{thm}
\label{p4-thm:para}
Assume that conditions (C1)-(C4) and (C9)-(C12) in Section \ref{p3-sec:proofs} hold. Then, as $n\rightarrow\infty$, $nh^4\rightarrow 0$, and $nh^2/\log(1/h)\rightarrow\infty$,
\begin{equation}
\sqrt{n}(\hat{\blambda}-\blambda)\overset{D}{\rightarrow} N(0,\bQ_2^{-1}),\notag
\end{equation}
where, \\
$\bQ_2=E\left[q_{\pi\pi}(Z)\begin{pmatrix}\bx\bpi'(Z)\\\textbf{I}\end{pmatrix}
\begin{pmatrix}\bx\bpi'(Z)\\\textbf{I}\end{pmatrix}^T-q_{\pi\pi}(Z)\begin{pmatrix}\bx\bpi'(Z)\\\textbf{I}\end{pmatrix}
\begin{pmatrix}\mathcal{I}_\pi^{(2)-1}(Z)E\{q_{\pi\pi}(Z)(\bx\bpi'(Z))^T|Z\}\\
\mathcal{I}_\pi^{(2)-1}(Z)E\{q_{\pi\eta}(Z)|Z\}\end{pmatrix}^T\right]$.
\end{thm}

\section{Numerical Studies}
\label{sec:p3-s4}
\subsection{Simulation Study}
In this section, we conduct simulation studies to test the performance of the proposed methodologies.

The performance of the estimates of the mean functions $m_j(\cdot)$'s in model (\ref{p3-model}) is measured by the square root of the average square errors (RASE)
\begin{equation}
RASE_m^2=N^{-1}\sum_{j=1}^k\sum_{t=1}^N[\hat{m}_j(u_t)-m_j(u_t)]^2.\notag
\end{equation}
In this simulation, we set $N=100$, and take the grid points are on the range. Similarly, we can define the $RASE$ for the variance functions $\sigma_j^2(\cdot)$'s and proportion functions $\pi_j(\cdot)$'s, denoted by $RASE_{\sigma^2}$ and $RASE_\pi$, respectively.

To apply the proposed methodologies, we use cross-validation (CV) to select a proper bandwidth for estimating the nonparametric functions.

\subsubsection{Example 1.}
We conduct a simulation for a 2-component MSIM:
\begin{center}
$\pi_1(z)=0.5+0.3\sin(\pi z)$ and $\pi_2(z)=1-\pi_1(z)$,\\
$m_1(z)=3-\sin(2\pi z/\sqrt{3})$ and $m_2(z)=\cos(\sqrt{3}\pi z)$,\\
$\sigma_1(z)=0.7+\sin(3\pi z)/15$ and $\sigma_2(z)=0.3+\cos(1.3\pi z)/10$.
\end{center}
where $z_i=\balpha^T\bx_i$, $\bx_i$ are trivariate with independent uniform (0,1) components, and the direction parameter is $\balpha=(1,1,1)/\sqrt{3}$. The sample sizes $n=200$, $n=400$, and $n=800$ are conducted over $500$ repetitions. To estimate $\balpha$, we use sliced inverse regression (SIR) and the fully iterative backfitting estimate (FIB). To estimate the nonparametric functions, we apply the one-step estimate (OS) and FIB. For FIB, we use both true value (T) and SIR (S) as the initial values.

We first select a proper bandwidth for estimating $\bpi(\cdot)$, $\bm(\cdot)$ and $\bsigma^2(\cdot)$. There are ways to calculate theoretical optimal bandwidth, but in practice, data driven methods, such as cross-validation (CV), are popularly used. Let $\mathscr{D}$ be the full data set, and divide $\mathscr{D}$ into a training set $\mathscr{R}_l$ and a test set $\mathscr{T}_l$. That is, $\mathscr{R}_l\cup \mathscr{T}_l=\mathscr{D}$ for $l=1,...,L$. We use the training set $\mathscr{R}_l$ to obtain the estimates $\{\hat{\bpi}(\cdot),\hat{\bm}(\cdot),\hat{\bsigma}^2(\cdot),\hat{\balpha}\}$. We then evaluate $\bpi(\cdot)$, $\bm(\cdot)$ and $\bsigma^2(\cdot)$ at the data in the corresponding training set. Then, for $(\bx_t,y_t)\in\mathscr{T}_l$, we calculate the classification probability as
\begin{equation}
\hat{p}_{tj}=\frac{\hat{\pi}_j(\hat{\balpha}^T\bx_t)\phi(y_t|\hat{m}_j(\hat{\balpha}^T\bx_t),\hat{\sigma}_j^2(\hat{\balpha}^T\bx_t))}{\sum_{j=1}^k\hat{\pi}_j(\hat{\balpha}^T\bx_t)\phi(y_t|\hat{m}_j(\hat{\balpha}^T\bx_t),\hat{\sigma}_j^2(\hat{\balpha}^T\bx_t))}
\end{equation}
for $j=1,...,k$. We then consider the regular $CV$, which is defined by
\begin{equation}
CV(h)=\sum_{l=1}^L\sum_{t\in \mathscr{T}_l}(y_t-\hat{y}_t)^2\notag,
\end{equation}
where $\hat{y}_t=\sum_{j=1}^k\hat{p}_{tj}\hat{m}_j(\hat{\balpha}^T\bx_t)$.

We set $L=10$ and randomly partition the data. We repeat the procedure 30 times, and take the average of the selected bandwidth as the optimal bandwidth, denoted by $\hat{h}$. In the simulation, we consider three different bandwidth, $\hat{h}\times n^{-2/15}$, $\hat{h}$ and $1.5\hat{h}$, which correspond to the under-smoothing, appropriate smoothing and over-smoothing condition, respectively.

Table \ref{p3-table1} reports the MSEs of $\hat{\balpha}$ (true value times 100). From Table \ref{p3-table1}, we can see that the fully iterative estimates give better results than SIR. We further notice that FIB(S) provides similar results to FIB(T), and therefore, SIR provides good initial values for other estimates.

Table \ref{p3-table2} contains the mean and standard deviation of $RASE_\pi$, $RASE_m$, and $RASE_{\sigma^2}$. We see that the fully iterative estimate is not sensitive to initial values.

\begin{table}[htb]
    \centering
\caption{MSE of $\hat{\balpha}$ (true value times 100)} \vskip 0.05in
\def\arraystretch{1.5}
\small \hspace*{-22.75pt}
\begin{tabular}{ c c|c| c c c | c c cc } \hline
\hline
&&SIR&&FIB(T)&&&FIB(S)&\\
\hline
&&&$h=0.054$&$h=0.109$&$h=0.164$&$h=0.054$&$h=0.109$&$h=0.164$\\
&$\alpha_1$&0.881&0.099&0.126&0.128&0.287&0.130&0.147\\
$n=200$&$\alpha_2$&0.829&0.113&0.144&0.124&0.324&0.144&0.137\\
&$\alpha_3$&1.066&0.110&0.152&0.137&0.388&0.154&0.167\\
\hline
&&&$h=0.045$&$h=0.100$&$h=0.149$&$h=0.045$&$h=0.100$&$h=0.149$\\
&$\alpha_1$&0.435&0.066&0.046&0.046&0.125&0.050&0.045\\
$n=400$&$\alpha_2$&0.447&0.063&0.054&0.051&0.121&0.055&0.052\\
&$\alpha_3$&0.411&0.062&0.052&0.052&0.123&0.053&0.052\\
\hline
&&&$h=0.037$&$h=0.091$&$h=0.137$&$h=0.037$&$h=0.091$&$h=0.137$\\
&$\alpha_1$&0.215&0.047&0.022&0.029&0.063&0.035&0.024\\
$n=800$&$\alpha_2$&0.256&0.034&0.035&0.040&0.044&0.029&0.027\\
&$\alpha_3$&0.226&0.065&0.031&0.058&0.062&0.050&0.030\\
\hline
\end{tabular}
\label{p3-table1}
\end{table}

\begin{table}[htb]
    \centering
\caption{Mean and Standard Deviation of RASEs} \vskip 0.05in
\def\arraystretch{1.5}
\small \hspace*{-22.75pt}
\begin{tabular}{ c c|c| c c c | c c cc } \hline
\hline
$n$&&OS&&FIB(T)&&&FIB(S)&\\
\hline
&&$h=0.125$&$h=0.054$&$h=0.109$&$h=0.164$&$h=0.054$&$h=0.109$&$h=0.164$\\
&$\pi$& 0.044(0.017)&0.057(0.015) &0.043(0.016)&0.049(0.017)&0.058(0.015) &0.043(0.016) &0.049(0.017)\\
200&$\mu$&0.227(0.063) &0.181(0.098) &0.176(0.046)&0.287(0.056)&0.178(0.086) &0.177(0.051) &0.288(0.059)\\
&$\sigma^2$&0.197(0.084) &0.175(0.169)&0.163(0.081)&0.246(0.071)&0.162(0.131)&0.164(0.095)&0.247(0.080)\\
\hline
&&$h=0.108$&$h=0.045$&$h=0.100$&$h=0.149$&$h=0.045$&$h=0.100$&$h=0.149$\\
&$\pi$&0.023(0.008)&0.032(0.008) &0.023(0.008)&0.027(0.009)&0.032(0.008) & 0.023(0.008) &0.027(0.009)\\
400&$\mu$&0.118(0.022)&0.093(0.045)&0.100(0.022) &0.169(0.020)&0.094(0.046) &0.100(0.022)&0.169(0.020)\\
&$\sigma^2$&0.104(0.035)& 0.089(0.077)&0.093(0.045)&0.143(0.028)&0.089(0.077)&0.093(0.045)&0.143(0.028)\\
\hline
&&$h=0.094$&$h=0.037$&$h=0.091$&$h=0.137$&$h=0.037$&$h=0.091$&$h=0.137$\\
&$\pi$&0.013(0.004)&0.017(0.003)&0.012(0.004)&0.016(0.004)&0.017(0.003)&0.012(0.004)&0.016(0.004)\\
800&$\mu$&0.062(0.010)&0.050(0.023)&0.056(0.010)&0.102(0.011)&0.050(0.023)&0.056(0.010)&0.101(0.010)\\
&$\sigma^2$&0.055(0.015)&0.049(0.046)&0.052(0.015)&0.086(0.010)&0.049(0.046)&0.051(0.012)&0.085(0.010)\\
\hline
\end{tabular}
\label{p3-table2}
\end{table}


\begin{figure}[htb]
    \centering
      \includegraphics[height=3in,width=5in]{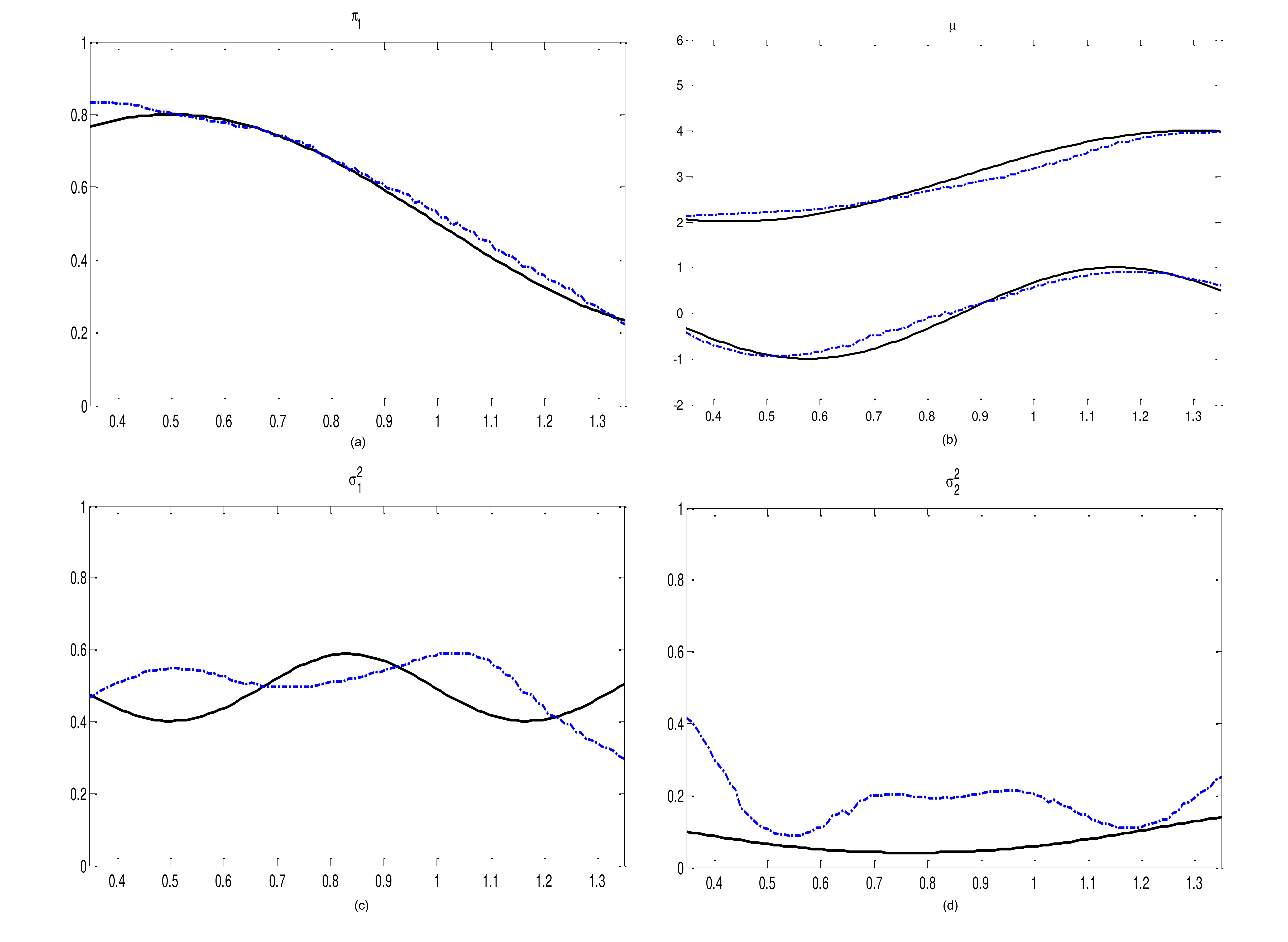}
    \caption{Simulation results of  for $n=800$ and $h=0.091$. The FIB(S) (dashed line) and true function (solid line) of: (a) $\pi_1$; (b) $\mu_1$ and $\mu_2$; (c) $\sigma_1^2$; and (d) $\sigma_2^2$.}
 \label{p3-figure2}
\end{figure}

\subsubsection{Example 2.}
We conduct a simulation for a 2-component MRSIP:
\begin{center}
$\pi_1(z)=0.5-0.35\sin(\pi z)$ and $\pi_2(z)=1-\pi_1(z)$,\\
$m_1(\bx)=1+3x_2$ and $m_2(\bx)=-1+2x_1+3x_3$,\\
$\sigma^2_1=0.7$ and $\sigma^2_2=0.6$,
\end{center}
where $m_1(\bx)$ and $m_2(\bx)$ are the regression functions for the first and second components, respectively. Therefore, $\bbeta_1=(1,0,3,0)$ and $\bbeta_2=(-1,2,0,3).$ $\bx_i$ are trivariate with independent uniform (0,1) components, and the direction parameter is $\balpha=(1,1,1)/\sqrt{3}$. MRSIP with true value (T) and SIR (S) as initial values are used to fit the data, and the results are compared to a two-component mixture of linear regression models (MixLinReg).

Table \ref{p41-tab1} reports the MSEs of parameter estimates, and Table \ref{p41-tab2} contains the MSEs of $\hat{\balpha}$ and the average of $RASE_\pi$. From both tables, we can see that MRSIP works comparable to MixLinReg when the sample size is small, and outperforms MixLinReg when sample size is big. We further notice that MRSIP(S) provides similar results to MRSIP(T), implying that SIR provides good initial values for MRSIP.

\begin{table}[htb]
    \centering
\caption{The MSEs of parameters (true value times 100)} \vskip 0.05in
\def\arraystretch{1.5}
\small \hspace*{-22.75pt}
\begin{tabular}{c c| c c c c c c c c | c cc } \hline
& & $\beta_{10}$ & $\beta_{11}$ & $\beta_{12}$ & $\beta_{13}$ & $\beta_{20}$ & $\beta_{21}$ & $\beta_{22}$ & $\beta_{23}$ & $\sigma^2_{1}$ & $\sigma^2_{2}$\\\hline
$n=200$ & MRSIP(S) & 46.37&32.78&34.73&37.61&11.19&16.55&15.05&16.36&4.649&1.754\\
\cline{2-12}
&MRSIP(T)& 51.91&33.62&39.01&37.25&11.10&16.56&15.07&16.04&4.584&1.649\\
\cline{2-12}
$h=0.131$& MixLinReg & 50.87&33.67&42.53&34.68&12.03&12.66&18.84&12.30& 4.250&1.265\\
\hline
$n=400$ & MRSIP(S) & 13.83&11.89&14.19&11.47&5.541&6.332&6.767&7.165&1.631&0.721\\
\cline{2-12}
&MRSIP(T)& 14.79&12.49&14.84&11.59&5.513&6.254&6.632&6.926&1.672&0.675\\
\cline{2-12}
$h=0.103$& MixLinReg & 29.03&14.97&29.46&15.72&8.045&5.967&12.46&6.269&1.864&0.626\\
\hline
$n=800$ & MRSIP(S) &  6.324&4.491&6.150&4.736&2.365&2.973&2.773&3.584&0.669&0.334\\
\cline{2-12}
&MRSIP(T)& 6.788&4.614&6.820&4.922&2.301&2.829&2.718&3.348&0.691&0.307\\
\cline{2-12}
$h=0.080$& MixLinReg & 21.89&6.866&21.84&8.223&5.413&3.163&8.775&3.640&0.848&0.352\\
\hline
\end{tabular}
\label{p41-tab1}
\end{table}

\begin{table}[htb]
    \centering
\caption{The MSEs of direction parameter and the average of RASE$_\pi$ (true value times 100)} \vskip 0.05in
\def\arraystretch{1.5}
\small \hspace*{-22.75pt}
\begin{tabular}{c c | c c c | cc } \hline
&&$\alpha_1$&$\alpha_2$&$\alpha_3$&RASE$_\pi$\\\hline
$n=200$&MRSIP(S)&  5.709&19.30&5.996&18.87\\
\cline{2-7}
&MRSIP(T)&   4.984&9.449&4.896&17.86\\
\cline{2-7}
$h=0.131$&MixLinReg&-&-&-&   28.98\\
\hline
$n=400$&MRSIP(S)&   2.682&6.968&3.029&13.74\\
\cline{2-7}
&MRSIP(T)&   2.113&3.019&1.902&12.98\\
\cline{2-7}
$h=0.103$&MixLinReg&-&-&-&    28.23\\
\hline
$n=800$&MRSIP(S)&   0.980&2.527&1.585&10.35\\
\cline{2-7}
&MRSIP(T)&    0.892&0.979&0.969&9.960\\
\cline{2-7}
$h=0.080$&MixLinReg&-&-&-&     28.04&\\
\hline
\end{tabular}
\label{p41-tab2}
\end{table}

\section{Real Data Example}
We illustrate the proposed methodology by an analysis of ``The effectiveness of National Basketball Association guards''.
There are many ways to measure the (statistical) performance of guards in the National Basket Association (NBA). Of interest is how the height of the player (Height), minutes per game (MPG) and free throw percentage (FTP) affects points per game (PPM) (Chatterjee et al., 1995).

\begin{figure}[htb]
    \centering
      \includegraphics[height=2in,width=7in]{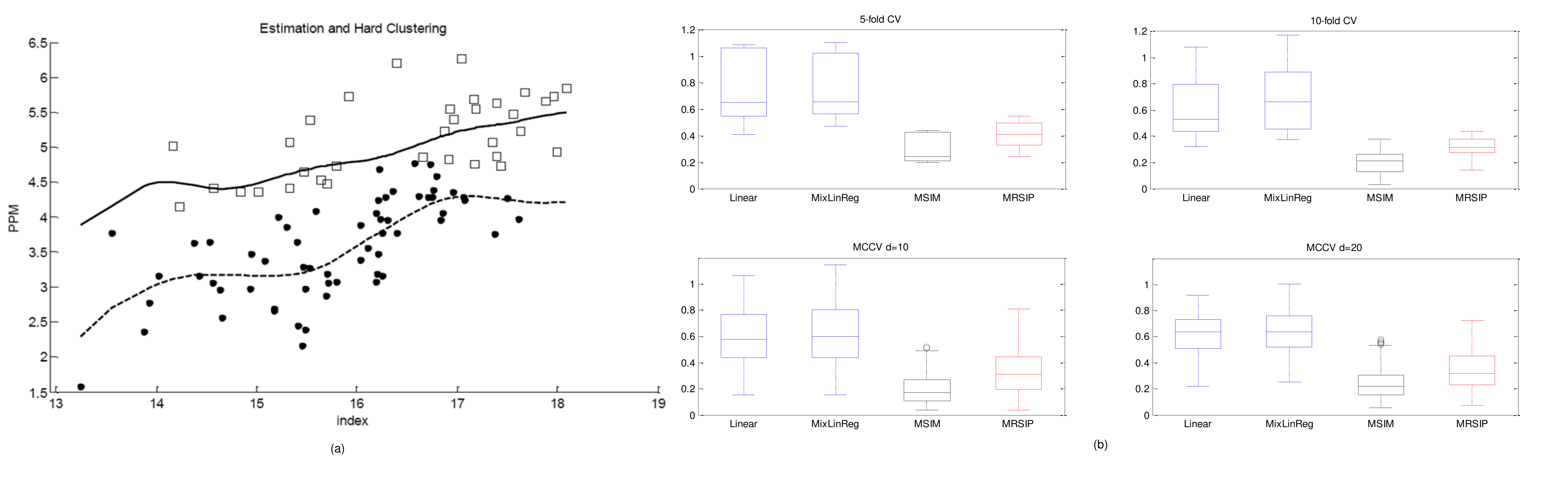}
    \caption{NBA data: (a) Estimated mean functions and a hard-clustering result; (b) Prediction accuracy: 5-fold CV; 10-fold CV; MCCV d=10; MCCV d=20.}
 \label{p3-figure3}
\end{figure}

The data set contains some descriptive statistics for all 105 guards for the 1992-1993 season. Since players playing very few minutes are quite different from those who play a sizable part of the season, we only look at those players playing 10 or more minutes per game and appearing in 10 or more games. We see that Michael Jordan is an outlier in terms of PPM, so we will also omit him from the data (Chatterjee et al., 1995). These excludes 10 players. We divide each variable by its corresponding standard deviation, so that they have comparable numerical scale. An optimal bandwidth is selected at 0.344 by CV procedure. Figure \ref{p3-figure3}(a) contains the estimated mean functions and hard-clustering results, denoted by dots and squares, respectively. The 95\% confidence interval for $\hat{\balpha}$ based on MSIM are (0.134,0.541), (0.715,0.949) and (0.202,0.679), indicating that MPG is the most influential factor on PPM.

To evaluate the prediction performance of the proposed model and compared it to linear regression model and mixture of linear regression models, we used $d$-fold cross-validation with $d$=5,10, and also Monte-Carlo cross-validation (MCCV) (Shao, 1993). In MCCV, the data were partitioned 500 times into disjoint training subsets (with size $n-d$) and test subsets (with size $d$). The mean squared prediction error evaluated at the test data sets over 500 replications are reported as boxplots in Figure \ref{p3-figure3}(b). Apparently, the MSIM and the MRSIP have superior prediction power than the linear regression model or the mixture of linear regression models, and MSIM is more favorable then the MRSIP for this data set.

\section{Discussion}
In this paper we proposed two finite semiparametric mixture of regression models and the corresponding backfitting estimates. We showed that the nonparametric functions can be estimated with the same rate as if the parameters were known and the parameters can be estimated with root-$n$ convergence rate. In this article, we assume that the number of components is known and fixed, but it requires more research to select the number of components for the proposed semiparametric mixture models. In addition, it is also interesting to build some formal test to compare the proposed two semiparametric mixture models. One way is to apply generalized likelihood ratio statistic proposed by Fan et al., (2001).

\section{Proofs}
\label{p3-sec:proofs}.
\noindent{\bf Technical Conditions:}
\begin{enumerate}[({C}1)]
\item The sample $\{(\bx_i,Y_i),i=1,...,n\}$ is independent and identically distributed from its population $(\bx,Y)$. The support for $\bx$, denoted by $\mathcal{X}$, is a compact subset of $\mathbb{R}^3$.
\item The marginal density of $\balpha^T\bx$, denoted by $f(\cdot)$, is twice continuously differentiable and positive at the point $z$.
\item The kernel function $K(\cdot)$ has a bounded support, and satisfies that
\begin{equation}
\int K(t)dt=1,\hspace{10 mm}\int tK(t)dt=0,\hspace{10 mm}\int t^2K(t)dt<\infty,\notag
\end{equation}
\begin{equation}
\int K^2(t)dt<\infty,\hspace{10mm}\int|K^3(t)|dt<\infty.\notag
\end{equation}
\item $h\rightarrow0$, $nh\rightarrow 0$, and $nh^5=O(1)$ as $n\rightarrow\infty$.
\item The third derivative $|\partial^3\ell(\btheta,y)/\partial\theta_i\partial\theta_j\partial\theta_k|\leq M(y)$ for all $y$ and all $\btheta$ in a neighborhood of $\btheta(z)$, and $E[M(y)]<\infty$.
\item The unknown functions $\btheta(z)$ have continuous second derivative. For $j=1,...,k$, $\sigma_j^2(z)>0$, and $\pi_j(z)>0$ for all $\bx\in\mathcal{X}$.
\item For all $i$ and $j$, the following conditions hold:
\begin{equation}
E\left[\left|\frac{\partial\ell(\btheta(z),Y)}{\partial\theta_i}\right|^3\right]<\infty\hspace{10 mm}
E\left[\left(\frac{\partial^2\ell(\btheta(z),Y)}{\partial\theta_i\partial\theta_j}\right)^2\right]<\infty\notag
\end{equation}

\item $\btheta_0''(\cdot)$ is continuous at the point $z$.
\item The third derivative $|\partial^3\ell(\bpi,y)/\partial\pi_i\partial\pi_j\partial\pi_k|\leq M(y)$ for all $y$ and all $\bpi$ in a neighborhood of $\bpi(z)$, and $E[M(y)]<\infty$.
\item The unknown functions $\bpi(z)$ have continuous second derivative. For $j=1,...,k$, $\pi_j(z)>0$ for all $\bx\in\mathcal{X}$.
\item For all $i$ and $j$, the following conditions hold:
\begin{equation}
E\left[\left|\frac{\partial\ell(\bpi(z),Y)}{\partial\pi_i}\right|^3\right]<\infty\hspace{10 mm}
E\left[\left(\frac{\partial^2\ell(\bpi(z),Y)}{\partial\pi_i\partial\pi_j}\right)^2\right]<\infty\notag
\end{equation}
\item $\bpi''(\cdot)$ is continuous at the point $z$.
\end{enumerate}

$\textbf{Proof of Theorem \ref{p3-thm:iden}.}$\\
Ichimura (1993) have shown that under conditions (i)-(iv), $\balpha$ is identifiable. Further, Huang et al. (2013) showed that with condition (v), the nonparametric functions are identifiable. Thus completes the proof.
\qed

$\textbf{Proof of Theorem \ref{p3-thm:nonp}.}$\\
Let\begin{align}
&\hat{\pi}_j^*=\sqrt{nh}\{\hat{\pi}_j-\pi_{j}(z)\},\hspace{3mm}j=1,...,k-1.\notag\\
&\hat{m}_j^*=\sqrt{nh}\{\hat{m}_j-m_{j}(z)\},\hspace{3mm}j=1,...,k,\notag\\
&\hat{\sigma}_j^{2*}=\sqrt{nh}\{\hat{\sigma}_j^2-\sigma_{j}^2(z)\},\hspace{3mm}j=1,...,k,\notag
\end{align}
Define $\hat{\bpi}^*=(\hat{\pi}_1^*,...,\hat{\pi}_{k-1}^*)^T$, $\hat{\bm}^*=(\hat{m}_1^*,...,\hat{m}_k^*)^T$, $\hat{\bsigma}^*=(\hat{\sigma}_1^*,...,\hat{\sigma}_k^*)^T$ and denote $\hat{\btheta}^*=(\hat{\bpi}^{*T},\hat{\bm}^{*T},(\hat{\bsigma}^{*2})^T)^T$. Let $a_n=(nh)^{-1/2}$. Let
\begin{align}
\ell(\btheta(z),\hat{\balpha},\bx_i,Y_i)=\log\left\{\sum_{j=1}^k\pi_j(\hat{\balpha}^T\bx_i)\phi(Y_i|m_j(\hat{\balpha}^T\bx_i),\sigma_j^2(\hat{\balpha}^T\bx_i))\right\}K_h(\hat{\balpha}^T\bx_i-z)\notag
\end{align}
If $(\hat{\bpi},\hat{\bm},\hat{\bsigma}^2)^T$ maximizes (\ref{p3-l1}), then $\hat{\btheta}^*$ maximizes
\begin{equation}
\ell_n^*(\btheta^*)=h\sum_{i=1}^n[\ell(\btheta(z)+a_n\btheta^*,\hat{\balpha},\bx_i,Y_i)-\ell(\btheta(z),\hat{\balpha},\bx_i,Y_i)]K_h(\hat{Z}_i-z)
\end{equation}
with respect to $\btheta^*$. By a Taylor expansion,
\begin{align}
\ell_n^*(\btheta^*)=\bW_{1n}^T\btheta^*+\frac{1}{2}\btheta^{*T}\bA_{1n}\btheta^*+o_p(1),
\end{align}
where
\begin{equation}
\bW_{1n}=\sqrt{\frac{h}{n}}\sum_{i=1}^n\frac{\partial\ell(\btheta(z),\hat{\balpha},\bx_i,Y_i)}{\partial\btheta}K_h(\hat{Z}_i-z),\notag
\end{equation}
and
\begin{equation}
\bA_{2n}=\frac{1}{n}\sum_{i=1}^n\frac{\partial^2\ell(\btheta(z),\hat{\balpha},\bx_i,Y_i)}{\partial\btheta\partial\btheta^T}K_h(\hat{Z}_i-z),\notag
\end{equation}
By WLLN, it can be shown that $\bA_{1n}=-f(z)\mathcal{I}^{(1)}_\theta(z)+o_p(1)$. Therefore,
\begin{align}
\ell_n^*(\btheta^*)=\bW_{1n}^T\btheta^*-\frac{1}{2}f(z)\btheta^{*T}\mathcal{I}^{(1)}_\theta(z)\btheta^*+o_p(1).
\end{align}
Using the quadratic approximation lemma (see, for example, Fan and Gijbels (1996)), we have that
\begin{equation}
\hat{\btheta}^*=f(z)^{-1}\mathcal{I}^{(1)}_\theta(z)^{-1}\bW_{1n}+o_p(1).
\end{equation}
Note that
\begin{align}
\bW_{1n}&=\sqrt{\frac{h}{n}}\sum_{i=1}^n\frac{\partial\ell(\btheta(z),\balpha,\bx_i,Y_i)}{\partial\btheta}K_h(Z_i-z)+D_{1n}+O_p(\sqrt{\frac{h}{n}}\|\hat{\balpha}-\balpha\|^2)\notag
\end{align}
where
\begin{align}
D_{1n}&=\sqrt{\frac{h}{n}}\sum_{i=1}^n\left\{\frac{\partial^2\ell(\btheta(z),\balpha,\bx_i,Y_i)}{\partial\btheta\partial\btheta^T}[\bx_i\btheta'(Z_i)]^TK_h(Z_i-z)\right\}(\hat{\balpha}-\balpha).\notag
\end{align}
Since $\sqrt{n}(\hat{\balpha}-\balpha)=O_p(1)$, it can be shown that $D_{1n}=-\sqrt{h}f(z)E[\frac{\partial^2\ell(\btheta(z),\balpha,\bx,Y)}{\partial\btheta\partial\btheta^T}[\bx\btheta'(Z)]^T]=o_p(1)$, and $O_p(\sqrt{\frac{h}{n}}\|\hat{\balpha}-\balpha\|^2)=o_p(1)$.
Therefore,
\begin{align}
\bW_{1n}&=\sqrt{\frac{h}{n}}\sum_{i=1}^n\frac{\partial\ell(\btheta,\balpha,\bx_i,Y_i)}{\partial\btheta}K_h(Z_i-z)+o_p(1)\notag.
\end{align}
To complete the proof, we now calculate the mean and variance of $\bW_n$. Note that
\begin{align}
E(\bW_{1n})&=\sqrt{nh}E\left[E[\frac{\partial\ell(\btheta,\balpha,\bx_i,Y_i)}{\partial\btheta}K_h(Z_i-z)|Z=z_0]\right]\notag\\
&=\sqrt{nh}[\frac{1}{2}f(z)\Lambda_1^{''}(z|z)+f'(z)\Lambda_1^{'}(z|z)]\kappa_2h^2.
\end{align}
Similarly, we can show that $\text{Cov}(\bW_{1n})=f(z)\mathcal{I}^{(1)}_\theta(z)\nu_0+o_p(1)$, where $\kappa_l=\int t^lK(t)dt$ and $\nu_l=\int t^lK^2(t)dt$. The rest of the proof follows a standard argument.\qed

$\textbf{Proof of Theorem \ref{p3-thm:index}.}$\\
Denote $Z=\balpha^T\bx$ and $\hat{Z}=\hat{\balpha }^T\bx$. Let $\ell(\btheta(z),X,Y)=\log\sum_{j=1}^k\pi_j(z)\phi(Y|m_j(z),\sigma_j^2(z))$.
If $\hat{\btheta}(z_0;\hat{\balpha})$ maximizes (\ref{p3-l1}), then it solves
\begin{equation}
\b0=n^{-1}\sum_{i=1}^n\frac{\partial\ell(\hat{\btheta}(z_0;\hat{\balpha}),X_i,Y_i)}{\partial\btheta}K_h(\hat{Z}_i-z_0).\notag
\end{equation}
Apply a Taylor expansion and use the conditions on $h$, we obtain
\begin{align}
\b0&=n^{-1}\sum_{i=1}^nq_{1i}(Z_i)K_h(Z_i-z_0)+
n^{-1}\sum_{i=1}^n\left[q_{2i}(Z_i)K_h(Z_i-z_0)\right](\hat{\btheta}(z_0;\hat{\balpha})-\btheta(z_0))\notag\\
&+n^{-1}\sum_{i=1}^nq_{2i}(Z_i)[\bx_i\btheta'(Z_i)]^TK_h(Z_i-z_0)(\hat{\balpha}-\balpha)+o_p(n^{-1/2})+O_p(h^2)\notag
\end{align}
By similar argument as in the previous proof,
\begin{align}
\hat{\btheta}(z_0;\hat{\balpha})-\btheta(z_0)&=n^{-1}f^{-1}(z_0)\mathcal{I}^{(1)-1}_\theta(z_0)\sum_{i=1}^nq_{1i}(Z_i)K_h(Z_i-z_0)\notag\\
&-\mathcal{I}^{(1)-1}_\theta(z_0)E\{q_2(Z)[\bx\btheta'(Z)]^T|Z=z_0\}(\hat{\balpha}-\balpha)+o_p(n^{-1/2})
\label{p3-A11}
\end{align}
Note that
\begin{align}
\label{p3-A13}
\hat{\btheta}&(\hat{\balpha}^T\bx_i;\hat{\balpha})-\btheta(\balpha^T\bx_i)=\hat{\btheta}(\hat{\balpha}^T\bx_i;\hat{\balpha})
-\hat{\btheta}(\balpha^T\bx_i;\hat{\balpha})+\hat{\btheta}(\balpha^T\bx_i;\hat{\balpha})-\btheta(\balpha^T\bx_i)\notag\\
&=(\hat{\btheta}'(\balpha^T\bx_i;\hat{\balpha}))^T(\hat{\balpha}^T-\balpha^T)\bx_i+\hat{\btheta}(\balpha^T\bx_i;\hat{\balpha})-\btheta(\balpha_0^T\bx_i)+o_p(n^{-1/2})\notag\\
&=(\btheta'(\balpha^T\bx_i))^T(\hat{\balpha}^T-\balpha^T)\bx_i+\hat{\btheta}(\balpha^T\bx_i;\hat{\balpha})-\btheta(\balpha^T\bx_i)+o_p(n^{-1/2})
\end{align}
where the second part is handled by (\ref{p3-A11}).

Since $\hat{\balpha}$ maximizes (\ref{p3-l2}), it is the solution to
\begin{align}
\b0=\lambda\hat{\balpha}+n^{-1/2}\sum_{i=1}^n\bx_i\hat{\btheta}'(\hat{\balpha}^T\bx_i;\hat{\balpha})
\frac{\partial\ell(\hat{\btheta}(\hat{\balpha}^T\bx_i;\hat{\balpha}),X_i,Y_i)}{\partial\btheta},\notag
\end{align}
where $\lambda$ is the Lagrange multiplier. By the Taylor expansion and using (\ref{p3-A13}), we have that
\begin{align}
\b0&=\lambda\hat{\balpha}+n^{-1/2}\sum_{i=1}^n \bx_i\btheta'(Z_i)q_{1i}(Z_i)+n^{-1/2}\sum_{i=1}^n \bx_i\btheta'(Z_i) q_{2i}(Z_i)
[\hat{\btheta}(\hat{\balpha}^T\bx_i)-\btheta(\balpha^T\bx_i)]+o_p(1)\notag\\
&=\lambda\hat{\balpha}+n^{-1/2}\sum_{i=1}^n\bx_i\btheta'(Z_i)q_{1i}(Z_i)+n^{-1/2}\sum_{i=1}^n \bx_i\btheta'(Z_i)
q_{2i}(Z_i)(\bx_i\btheta'(Z_i))^T(\hat{\balpha}-\balpha)\notag\\
&+n^{-1/2}\sum_{i=1}^n\bx_i\btheta'(Z_i)q_{2i}(Z_i)
[\hat{\btheta}(Z_i)-\btheta(Z_i)])+o_p(1).\notag
\end{align}
Define
\begin{equation}
A_\alpha=E\{[\bx\btheta'(Z)]q_2(Z)[\bx\btheta'(Z)]^T\},\notag
\end{equation}
and apply (\ref{p3-A11}),
\begin{align}
\label{p3-A14}
\b0&=\lambda\hat{\balpha}+n^{-1/2}\sum_{i=1}^n\bx_i\btheta'(Z_i)
q_{1i}(Z_i)+n^{1/2}A_\beta(\hat{\balpha}-\balpha)\notag\\
&-n^{-1/2}\sum_{i=1}^n\bx_i\btheta'(Z_i)q_{2i}(Z_i)\mathcal{I}^{-1}_\theta(Z_i)E\{q_2(Z)[\bx\btheta'(Z)]^T|Z=Z_i\}(\hat{\balpha}-\balpha)\notag\\
&+n^{-1/2}\sum_{i=1}^n\bx_i\btheta'(Z_i)q_{2i}(Z_i)
n^{-1}f^{-1}(Z_i)\mathcal{I}_\theta^{-1}(Z_i)\sum_{t=1}^nq_{1t}(Z_t)K_h(Z_t-Z_i)+o_p(1)\notag\\
&=\lambda\hat{\balpha}+n^{-1/2}\sum_{i=1}^n\bx_i\btheta'(Z_i)q_{1i}(Z_i)
+\bQ_1 n^{1/2}(\hat{\balpha}-\balpha)\notag\\
&+n^{-1/2}\sum_{i=1}^n\bx_i\btheta'(Z_i)q_{2i}(Z_i)
n^{-1}f^{-1}(Z_i)\mathcal{I}^{(1)-1}_\theta(Z_i)\sum_{t=1}^nq_{1t}(Z_t)K_h(Z_t-Z_i)+o_p(1).
\end{align}
Interchanging the summations in the last term, we get
\begin{align}
\label{p3-A15}
&n^{-1/2}\sum_{i=1}^n\left[n^{-1}\sum_{t=1}^n\bx_t\btheta'(Z_t)q_{2t}(Z_t)
K_h(Z_t-Z_i)f^{-1}(Z_t)\mathcal{I}_\theta^{-1}(Z_t)q_{1i}(Z_i)\right]\notag\\
&=n^{-1/2}\sum_{i=1}^nE[\bx\btheta'(Z)q_2(Z)|Z_i]\mathcal{I}^{(1)-1}_\theta(Z_i)q_{1i}(Z_i)+o_p(1)
\end{align}

Let $\Gamma_\alpha=I-\balpha\balpha^T+o_p(1)$. Combining (\ref{p3-A14}) and (\ref{p3-A15}), and multiply by $\Gamma_\alpha$, we have
\begin{align}
\label{p3-A12}
\Gamma_\alpha\bQ_1 n^{1/2}(\hat{\balpha}-\balpha)=n^{-1/2}\sum_{i=1}^n\Gamma_\alpha\{\bx_i\btheta'(Z_i)+E[\bx\btheta'(Z)q_2(Z)|Z_i]\mathcal{I}^{(1)-1}_\theta
(Z_i)\}q_{1i}(Z_i)+o_p(1)
\end{align}
It can be shown that the right-hand side of (\ref{p3-A12}) has the covariance matrix $\Gamma_\alpha\bQ_1\Gamma_\alpha$, and therefore, completes the proof.\qed

$\textbf{Proof of Theorem \ref{p4-thm:iden}}$\\
Ichimura (1993) have shown that under conditions (i)-(iv), $\balpha$ is identifiable. Furthermore, Huang and Yao (2012) showed that with condition (v), $(\bpi(\cdot),\bbeta,\bsigma^2)$ are identifiable. Thus completes the proof.\qed

$\textbf{Proof of Theorem \ref{p4-thm:nonp}}$\\
This proof is similar to the proof of \ref{p3-thm:nonp}

Let $\hat{\pi}_j^*=\sqrt{nh}\{\hat{\pi}_j-\pi_j(z)\}$, $j=1,...,k-1$, and $\hat{\bpi}^*=(\hat{\pi}_1^*,...,\hat{\pi}_{k-1}^*)^T$. It can be shown that
\begin{equation}
\hat{\bpi}^*=f(z)^{-1}\mathcal{I}_\pi^{(2)-1}(z)\bW_{2n}+o_p(1).\notag
\end{equation}
where
\begin{equation}
\bW_{2n}=\sqrt{\frac{h}{n}}\sum_{i=1}^n\frac{\partial\ell(\bpi(z),\hat{\blambda},\bx_i,Y_i)}{\partial\bpi}K_h(\hat{Z}_i-z).\notag
\end{equation}
To complete the proof, notice that
\begin{align}
E(\bW_{2n})=&\sqrt{nh}E\left\{E[\frac{\partial\ell(\bpi,\blambda,\bx_i,Y_i)}{\partial\bpi}K_h(Z_i-z)|Z=z_0]\right\}\notag\\
=&\sqrt{nh}[\frac{1}{2}f(z)\Lambda_2''(z|z)+f'(z)\Lambda_2'(z|z)]\kappa_2h^2.\notag
\end{align}
and Cov$(\bW_{2n})=f(z)\mathcal{I}^{(2)}_\pi(z)\nu_0+o_p(1)$. The rest of the proof follows a standard argument.\qed\\

$\textbf{Proof of Theorem \ref{p4-thm:para}}$\\
The proof is similar to the proof of Theorem \ref{p3-thm:index}.
It can be shown that
\begin{align}
&\hat{\bpi}(z_0;\hat{\blambda})-\bpi(z_0)=n^{-1}f^{-1}(z_0)\mathcal{I}^{(2)-1}_\pi(z_0)\sum_{i=1}^nq_{\pi i}(Z_i)K_h(Z_i-z_0)\notag\\
-&\mathcal{I}^{(2)-1}_\pi(z_0) E\{q_{\pi\pi}(Z)[\bx\bpi'(Z)]^T|Z=z_0\}(\hat{\balpha}-\balpha)-\mathcal{I}^{(2)-1}_\pi(z_0) E\{q_{\pi\eta}(Z)|Z=z_0\}(\hat{\bta}-\bta)+o_p(n^{-1/2}),\notag
\label{p4-e1}
\end{align}
and therefore,
\begin{equation}
\hat{\bpi}(\hat{Z}_i;\hat{\blambda})-\bpi(Z_i)=\{\bx_i\bpi'(Z_i)\}^T(\hat{\balpha}-\balpha)+\hat{\bpi}(Z_i;\hat{\blambda})-\bpi(Z_i)+o_p(n^{-\frac{1}{2}}).
\label{p4-e2}
\end{equation}
Since $\hat{\blambda}$ maximizes (\ref{p41-l2}), it is the solution to
\begin{align}
\b0=\gamma\begin{pmatrix}\hat{\balpha}\\\b0\end{pmatrix}+n^{-\frac{1}{2}}\sum_{i=1}^n\begin{pmatrix}\bx_i\hat{\bpi}'(\hat{Z}_i;\hat{\blambda})\\
\textbf{I}\end{pmatrix}q_\pi(\hat{\bpi}(\hat{Z}_i;\hat{\blambda}),\hat{\blambda}),\notag
\end{align}
where $\gamma$ is the Lagrange multiplier. By Taylor series and (\ref{p4-e2})
\begin{align}
\b0=&\gamma\begin{pmatrix}\hat{\balpha}\\\b0\end{pmatrix}+n^{-\frac{1}{2}}\sum_{i=1}^n\bLambda_{1i}q_{\pi i}(Z_i)+n^{\frac{1}{2}}\bQ_2\begin{pmatrix}\hat{\balpha}-\balpha\\\hat{\bta}-\bta\end{pmatrix}\notag\\
+&n^{-\frac{1}{2}}\sum_{i=1}^n\bLambda_{1i}q_{\pi\pi i}(Z_i)n^{-1}f^{-1}(Z_i)\mathcal{I}^{(2)-1}_\pi(Z_i)\sum_{j=1}^nq_{\pi j}(Z_j)K_h(Z_j-Z_i)+o_p(1)\notag\\
=&\gamma\begin{pmatrix}\hat{\balpha}\\\b0\end{pmatrix}+n^{-\frac{1}{2}}\sum_{i=1}^n\bLambda_{1i}q_{\pi i}(Z_i)+n^{\frac{1}{2}}\bQ_2\begin{pmatrix}\hat{\balpha}-\balpha\\\hat{\bta}-\bta\end{pmatrix}\notag\\
+&n^{-\frac{1}{2}}\sum_{i=1}^nE[\bLambda_{1i}q_{\pi\pi}(Z_i)]\mathcal{I}^{(2)-1}_\pi(Z_i)q_{\pi i}(Z_i)+o_p(1).
\label{p4-e3}
\end{align}
where $\bLambda_{1i}=\begin{pmatrix}\bx_i\bpi'(Z_i)\\\textbf{I}\end{pmatrix}$, and the last equation is the result of interchanging the summations.
Let $\Gamma_\alpha=\begin{pmatrix}\bI-\balpha\balpha^T&\textbf{0}\\\textbf{0}&\bI\end{pmatrix}+o_p(1)$. By (\ref{p4-e3}), and multiply by $\Gamma_\alpha$, we have
\begin{equation}
n^{\frac{1}{2}}\Gamma_\alpha\bQ_2\begin{pmatrix}\hat{\balpha}-\balpha\\\hat{\bta}-\bta\end{pmatrix}=n^{-\frac{1}{2}}\sum_{i=1}^n\Gamma_\alpha
\left\{\bLambda_{1i}-\mathcal{I}_\pi^{(2)-1}(Z_i)E[\bLambda_{1i}(Z_i)q_{\pi\pi}(Z_i)|Z_i]\right\}q_{\pi i}(Z_i)+o_p(1).
\label{p4-e5}
\end{equation}
It can be shown that the right-hand side of (\ref{p4-e5}) has the covariance matrix $\Gamma_\alpha\bQ_2\Gamma_\alpha$, and thus, completes the proof.\qed

\end{document}